\pdfoutput=1
\UseRawInputEncoding
\documentclass[%
 reprint,
 amsmath,amssymb,
 aps,
]{revtex4-2}

\usepackage{graphicx}% Include figure files
\usepackage{dcolumn}% Align table columns on decimal point
\usepackage{subfig}
\usepackage{bm}% bold math
\usepackage {float}
\begin{document}

%\preprint{APS/123-QED}

\title{Studies on dark energy evolution}% Force line breaks with \\
%\thanks{A footnote to the article title}%

\author{Daniela Grandón$^{1}$}
\email{daniela.grandon@ug.uchile.cl}
\author{Victor H. Cardenas$^{2}$}%
 \email{victor.cardenas@uv.cl}

\affiliation{$^{1}$ Grupo de Cosmología y Astrofísica Teórica, Departamento de Física, FCFM, Universidad de Chile, Blanco Encalada 2008, Santiago, Chile}

\affiliation{$^{2}$ Instituto de Física y Astronomía, Universidad de Valparaíso, Av. Gran Bretana 1111, Valparaíso, Chile}

\date{\today}% It is always \today, today,
             %  but any date may be explicitly specified

\begin{abstract}
In this work we explore signatures of evolution for the dark energy density $X(z)=\rho_{de}(z)/\rho_{de}(0)$ using latest observations on SNIa and H(z). The models consist of parametrizations of the dark energy density and consequently a reconstruction for the EoS parameter w(z) as a function of redshift. Both parametrization methods using the SH0Es prior results in a small deviation from LCDM at 1$\sigma$ for $X(z)$. Extending the analysis up to 2$\sigma$, the evidence for evolution of $X(z)$ dilute in both cases. We have also studied an interacting dark model where this trend is also found.
\end{abstract}

%\keywords{Suggested keywords}%Use showkeys class option if keyword
                              %display desired
\maketitle

%\tableofcontents

\section{\label{sec:intro} Introduction}

Since the discovery in 1998  \cite{Riess:1998cb}, \cite{Perlmutter_1999} that high redshift supernovae were fainter than expected, the scientific community has had to accept the existence of a new component in the universe, which in its most simplest form is known as a cosmological constant of a very specific magnitude. 
%ok

Although simple, this is unsatisfactory in many ways. First of all, this constant value seems to be fine tuned in order to reproduce the effect observed in the supernovae magnitude. Actually this value is of many orders of magnitude off the value suggested from particle physics. Second, there is no clue about the origin of this component, because is clear that no connection exist with the particle physics content in the universe, and third, the fact that being a constant during all the universe history makes today a very singular epoch in the cosmic evolution when this component becomes dominant, the so called {\it cosmic coincidence problem}.
%ok

In the context of the standard model of cosmology, and moving beyond the cosmological constant hypothesis, this can be explained in three ways: a new component filling the universe, as a quintessence field, a modification of the theory of gravity, and also relaxing the cosmological principle, considering inhomogeneous models. Given the level of ignorance about what is the actual cause of this effect, we generally call it dark energy (DE) \cite{Frieman:2008sn}, \cite{Huterer_2017}.
%ok

Because we do not have any idea about the nature of DE, most of the studies have followed a phenomenological approach. In this case, we proceed assuming a specific form for the EoS parameter $p_{de}/\rho_{de}=w(z)$. This is the idea behind the Chevalier-Polarski-Linder \cite{Chevallier:2000qy}, \cite{Linder:2002et} parametrization, for example, where its assumed that $w(z) = w_0+w_a(1-a)$, appropriated for quintessence models.
%ok

A particular interesting approach in this context is to follow the opposite direction: use the data to reconstruct the form of the cosmological functions, as $w(z)$. However, we faced a problem here. As was demonstrated in \cite{Maor_2001}, because the $w(z)$ function is behind two integrals from the data we use to reconstruct it, any features that may have $w(z)$ would be certainly erased after the analysis. For example, in \cite{Tegmark_2002} a reconstruction was made using $H(z)$ and the growth function, and the authors of \cite{Huterer:2002hy} propose a reconstruction of the scalar field potential assuming that a quintessence field is the nature of DE.
%ok

In this context, reconstruction schemes using the DE density $\rho_{de}$ were proposed by \cite{Wang_2002} and \cite{Wang:2001ht}. In \cite{Wang:2003gz},\cite{Wang:2004ru} the authors reconstruct the DE density assuming a parametrization for $X(z)=\rho_{de}/\rho_{de}^0$, showing better results that using the EoS parameter. There have been also works that attempts to reconstruct the EoS of DE using the Padé approximation \cite{mehrabi_2018}, non-parametric methods, such as Gaussian Processes (GP) \cite{Holsclaw:2010nb} and MCMC \cite{Holsclaw:2010sk} or Principal-Component approach \cite{Crittenden:2011aa}, where no functional form for DE density is assumed. Recently in \cite{Zhai:2018vmm} the DE density was reconstructed using a cubic spline in a set of redshifts, and also in \cite{Heisenberg:2020ywd} where a method of reconstruction based on the use of the shifted Chebyshev polynomials was used.
%ok

In most of these studies they agree that the use of priors on the Hubble constant parameter $H_0$ helps to reduce the uncertainties in the reconstruction.
%ok

In this work we use the energy density parameter as a probe to detect departures from the LCDM model. This can be considered as a continuation of our work in \cite{C_rdenas_2015} where we detect appreciable variations of $X (z)=\rho_{de}/\rho_{de}^0$ using data at low redshift. Here, we retake that analysis using the latest supernova data, the Pantheon sample \cite{Scolnic_2018} together with a larger sample of $H(z)$ measurements, and we have also add to the quadratic parametrization a cubic one, just to be sure about the trend obtained using only a quadratic model.
%ok

This work is organized as follows. In the next section we describe the parametrization methods and the main equations to be used in this paper. In section III we present the data samples and the characteristic $\chi^2$ for each of them. In section IV we present our results for both the quadratic and cubic parametrization, and in section V we discuss a theoretical model that show the trend evolution the data suggest. We finally end with our conclusions in section VI. 
%ok

\section{\label{sec:method} The method}

In this section we describe our methods for the subsequent analysis.

Because DE is a component whose nature is completely unknown, there is no clear point to postulate {\it a priori} what model should be choose or what fundamental idea is behind it. Assuming the level of ignorance we have about this new component, we decided to use the density of DE as a probe to explore the form it should take based on what the observational data suggests. Only after this work has been performed, we can have certain idea what model could explain the observed data best. So we step back and consider the DE density $X(z)$ as a probe to be constrained from observations. Our intention here is to detect departures from the LCDM model where this function is by definition equal to one at all time.
%ok

Close to this spirit is the well known use of the Chevallier-Polarski-Linder \cite{Chevallier:2000qy}, \cite{Linder:2002et} parametrization of the equation of state parameter $w(z) = w_0 + w_az/(1+z)$. There have been many studies using $w(z)$ as the probe function that after test it against observation is reconstructed. For example, in \cite{Zhao:2012aw} the authors using a non parametric method found that in fact $w(z)$ evolves crossing the phantom line at $z \simeq 0.25$. The problem in using $w(z)$ as a probe is that this function is two integrals apart from the observational data, causing the information to be diluted. This is clearly seen in the following expressions. The apparent magnitude measured of the supernovae depends directly on the luminosity distance defined as
\begin{equation}\label{lumdist}
d_L(z)=(1+z)\frac{c}{H_0}\int_{0}^{z}\frac{dz'}{E(z')},
\end{equation}
assuming a flat universe. The cosmology is encoded in the normalized Hubble function $E(z)=H(z)/H_0$ given by
\begin{equation}\label{edz}
E(z)^2 = \Omega_m (1+z)^3 + \Omega_r(1+z)^4 + (1-\Omega_m -\Omega_r)X(z),
\end{equation}
where here $X(z)$ is the normalized DE density, and its relation with the EoS parameter $w(z)$ is given through a new integral by
\begin{equation}\label{wdz}
 X(z) = \exp\left[ 3\int_{0}^z \frac{1+w(z')}{1+z'}dz'\right].
\end{equation}
This problem was pointed out immediately after the detection of accelerated expansion  \cite{Maor_2001} and the emergence of studies looking for evolution of $w(z)$. For this reason we focus on parametrizations on $X(z)$ only. Once we have reconstructed $X(z)$ we can re-derive the $w(z)$ -- inverting relation (\ref{wdz}) -- and even the deceleration parameter $q(z)$ can be derived from the best fit values.
%ok

\subsection{Quadratic parametrization}

Following our previous work \cite{C_rdenas_2015}, we use first a quadratic interpolation for $X(z)$ given by
\begin{eqnarray}\nonumber
X(z) &=&  \frac{x_0(z-z_1)(z-z_2)}{(z_0-z_1)(z_0-z_2)} + 
\frac{x_1(z-z_0)(z-z_2)}{(z_1-z_0)(z_1-z_2)} + \\ &+&\frac{x_2(z-z_0)(z-z_1)}{(z_2-z_0)(z_2-z_1)},\label{xdz1}
\end{eqnarray}
where $x_0$, $x_1$ and $x_2$ are the constant values of $X(z)$ evaluated at $z_0$, $z_1$ and $z_2$ where we assume here that $z_2>z_1>z_0$. Given the definition of $X(z)$ the value for $x_0$ is fixed at $z_0=0$ being $x_0=1$. For $z_2$ we use the largest redshift in the data $z_m$ that allows us to properly estimate the parameters and for the intermediate value we take $z_1=z_m/2$. Under these definitions the free parameters are: $x_1=X(z_m/2)$ and $x_2=X(z_m)$. In this case (\ref{xdz1}) reduces to
\begin{equation}\label{inter1}
X(z)= 1+\frac{z (4x_1 -x_2-3)}{z_m}-\frac{2 z^2 (2
   x_1-x_2-1)}{z_m^2}.
\end{equation}
If the data prefer the null result -- no evolution -- means that both parameters should prefer a value equal to unity.
%ok

\subsection{Cubic parametrization}

Knowing the limitations of a quadratic interpolation -- this function can only detect a single trend, either detect a parabolic increase or decrease with the evolution -- in this section we expand in a single parameter the previous analysis assuming a cubic parametrization. Of course it has the disadvantage of add a new free parameter in the fit compared to the quadratic parametrization (and two more respect to the cosmological constant). In this case we have 
\begin{eqnarray}\nonumber
X(z)=  \frac{x_0(z-z_1)(z-z_2)(z-z_3)}{(z_0-z_1)(z_0-z_2)(z_0-z_3)} + \\ \nonumber
+ \frac{x_1(z-z_0)(z-z_2)(z-z_3)}{(z_1-z_0)(z_1-z_2)(z_1-z_3)} +\\ \nonumber + \frac{x_2(z-z_0)(z-z_1)(z-z_3)}{(z_2-z_0)(z_2-z_1)(z_2-z_3)} + \\  + \frac{x_3(z-z_0)(z-z_1)(z-z_2)}{(z_3-z_0)(z_3-z_1)(z_3-z_2)} .\label{xdz2}
\end{eqnarray}
As we did before, we assume $z_0<z_1<z_2<z_3$ and set $z_0=0$ in such a way that $x_0=X(z=z_0=0)=1$, and we set $z_3=z_m$ as the largest redshift in the data sets that enable us to obtain a good fit of the parameters, and $z_2=2z_m/3$ and $z_1=z_m/3$. In this case the parameters $x_1=X(z_1)$, $x_2=X(z_2)$ and $x_3=X(z_3)$ are the free parameters to be constrained against the data. In this case (\ref{xdz2}) reduces to
\begin{eqnarray}\nonumber
X(z)= 1 + 9 (-1 + 3 x_1 - 3 x_2 + x_3) \frac{z^3}{2z_m^3} +\\ \nonumber
  - 9 (-2 + 5 x_1 - 4 x_2 + x_3)\frac{z^2}{2z_m^2} + \\ + (-11 + 18 x_1 - 9 x_2 + 
      2 x_3) \frac{z}{2z_m}
\end{eqnarray}
As before, the expectation for a LCDM model is that $x_1=x_2=x_3=1$.

\section{\label{sec: data} Data}

\subsection{Type Ia supernova}

For SNIa we use the Pantheon sample \cite{Scolnic_2018} consisting in 1048 data points which gives us the redshift, apparent magnitude at maximum brightness, and the covariance and correlations among the data. 
They span in the redshift range $0.01 < z <2.3$. We compute the residuals $\mu - \mu_{th}$ and minimize the quantity
\begin{equation}\label{chi2jla}
\chi^2 = (\mu - \mu_{th})^{T} C^{-1}(\mu - \mu_{th}),
\end{equation}
where $\mu_{th} = 5 \log_{10} \left( d_L(z)/10pc\right) $ gives the theoretical distance modulus, $d_L(z)$ is the luminosity distance, $ C $ is the covariance matrix released in \cite{Scolnic_2018}. Because the absolute magnitude and the Hubble parameter are degenerated in the computation of the distance modulus, we marginalize over these nuisance parameters using
\begin{equation}
    \chi^2_{sn} = A + \log \frac{D}{2\pi} - \frac{B^2}{D},
\end{equation}
where $A=(\mu - \mu_{th})^T C^{-1}(\mu - \mu_{th})$, $B = (\mu - \mu_{th})^{T} C^{-1} {\bf 1}$ and $D= {\bf 1}^{T} C^{-1} {\bf 1}$, as is explained in \cite{Conley_2010}.

\subsection{$H(z)$ measurements}

The data points for $H(z)$ can be obtained through two methods: the differential age (DA) method \cite{Jimenez:2001gg}, \cite{Simon_2005}, \cite{Stern_2010} which are based on
\begin{equation}
    H(z) = \frac{1}{a}\frac{da}{dt}=-\frac{1}{1+z}\frac{dz}{dt}\simeq -\frac{1}{1+z}\frac{\Delta z}{\Delta t},
\end{equation}
which so far consist in 31 measurements compiled from \cite{Wei:2018cov} spanning a redshift range of $(0.07,1.965)$. The second method is based on measurements of the line of sight BAO data \cite{Blake_2012}, \cite{Chuang_2013}, \cite{Font_Ribera_2014}, \cite{Delubac_2015}, \cite{Bautista_2017}, which consist in 26 data points extra. However, there is an overlap of three points between data from the WiggleZ and the BOSS team. We have excluded the three data points from \cite{Blake_2012}. In summary, for this observational probe we have $54$ data points which can constrain the Hubble function (\ref{edz}) through the $\chi^2$
\begin{equation}
    \chi_{hz}^2 = \sum_{i=1}^{54} \frac{\left\{ H_i -100 h E(z_i) \right\}^2}{\sigma_i^2},
\end{equation}
where $H_i$ are the values of the Hubble function at redshift $z_i$ measured with error $\sigma_i$.
%ok

The analysis is performed using a public code known as \texttt{emcee} \cite{Foreman_Mackey_2013}. This is a stable, well tested Python implementation of the affine-invariant ensemble sampler for Markov chain Monte Carlo (MCMC) proposed by Goodman \& Weare \cite{2010CAMCS...5...65G}. The output from the chains are visualized using GetDist \cite{Lewis:2019xzd}.

\section{\label{sec: results} Results}

In this section we describe the details of the analysis for each parametrization and also describe the main results. The procedure is the following.
%ok

We consider a cosmological model with standard dark matter and a dark energy component given by our parametrization (\ref{xdz1}) or (\ref{xdz2}) defining $E(z)$ as in (\ref{edz}). With this function we compute the joint $\chi^2=\chi_{sn}^2 + \chi_{hz}^2$ statistic defined in the previous section. We add one gaussian prior on the matter contribution, $\Omega_m = 0.315 \pm 0.021$ which is based on \cite{10.1093/ptep/ptaa104} but where we have increased three times the error to allow variations around the best fit value. Further we also add a prior on $H_0$. At this point, given the well-known debate on the $H_0$ tension we explore the effects of using two different gaussian priors on this parameter: using the Planck \cite{Aghanim:2018eyx} value $H_0= 67.5 \pm 1.5$ and using the SH0ES \cite{Riess:2019cxk} value $H_0=73.5 \pm 1.4$.
%ok

\subsection{Choosing $z_m$}
\begin{figure}[h!]
\centering
\includegraphics[width=8cm]{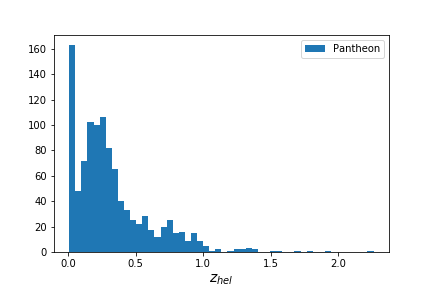}
\caption{Redshift distribution for Pantheon data} \label{fig: hist_sn}
\end{figure}

\begin{figure}[h!]
\centering
\includegraphics[width=8cm]{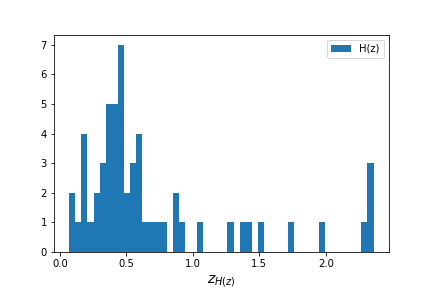}
\caption{Redshift distribution for $H(z)$ data} \label{fig: hist_hz}
\end{figure}

A key thing to discuss is the election of $z_m$. It is used in the parametrization (\ref{xdz1}) and (\ref{xdz2}), as the maximum redshift for the data set, however choosing a particular value has a direct impact in the best fit parameters and errors we get. First of all, let us see the redshift distribution in our data sets. For type Ia supernova we get the distribution displayed in Fig.(\ref{fig: hist_sn}). 
%\begin{figure}[h!]
%\centering
%\includegraphics[width=8cm]{errorbars_zhz.png}
%\caption{H(z) data} \label{fig1}
%\end{figure}
%
As can be seen, most of the data is below $z =0.5$ with fewer and fewer data for $z > 1.0$. Although there are data points for redshifts larger than $1.5$, we decided not to use them as the anchor for the larger redshift in the parametrization ($z_m$), although they are used in the fitting. The histogram for $H(z)$ measurements is shown in Fig.(\ref{fig: hist_hz}).
Being only 54 data points, the covering is sparse and most of the data is in the redshift range $[0, 0.7]$. Because we want to perform a joint analysis -- SNIa + $H(z)$ -- we study first the effect in the best fit parameters choosing different values for $z_m$ taken from $0.4$ until $1.5$ (beyond which, as we already mentioned, we do not expect to get sensitive feedback) with a step of $0.1$, performing several runs of our MCMC code. After that, we analyse these results and choose a value for $z_m$ to obtain the best fit parameters for each parametrization. Let us see in detail what all this part of the analysis is about. Here, the main question is which value of $z_m$ should be used and why. Given our goal is to reconstruct both the DE density and the EoS parameter $w(z)$, a potential problem arises: both parametrizations (5) and (7) are expansions in terms of powers of $z/z_m$, then by choosing a $z_m$ below the maximum redshift in the data used in the analysis, implies we will have data points in the reconstruction of $X(z)$ (those data with $z>z_m$) which will have an exaggerated contribution, distorting the evolution.
%ok

In order to overcome this potential problem, we carry out a detailed study of the effects on the choice of a particular value of $z_m$.%, emphasizing those that produces the less dispersion of errors in the best fit parameters.
The results of the MCMC chains with different values for $z_m$ are summarize in Figures (3), (4), (8) and (9), each one containing a specific combination between parametrization method and prior.%  and choose that with minimal errors.

We see that best fits parameters show an evolution as $z_m$ increases, and so do the error bars. As an example, in Fig.(3) the point $z_m$ equal to 1.2 produces a best fit for $x_2$ which clearly deviates from 1 towards smaller values, however, its error bars increases with respect to other $z_m$. So in conclusion, as long as $z_m$ increases, the error bars increases covering a wider range of values implying that we start loosing information from the fit. Despite this the statistical analysis shows the method is under control because the trend (that of getting lower values for the parameters as we increase $z_m$) remains the same. No instabilities were found after running several MCMC chains.%This occurs although the statistical analysis shows this is under control, because the trend (that of getting lower values for the parameters as we increase $z_m$) remains the same. No instabilities were found after running several MCMC chains.

Finally we use our results to reconstruct the $X(z)$ and the derived function from it: the EoS parameter $w(z)$. Because this analysis depends on which parametrization we use, we next separate the results in two sections each one for the parametrization (\ref{xdz1}) and (\ref{xdz2}).
%ok

\subsection{Quadratic case}

\begin{figure}[h!]
\centering
\includegraphics[width=8cm]{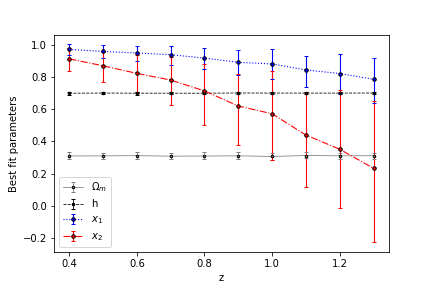}
\caption{Best fit values for $\Omega_m$, $h$, $x_1$ and $x_2$ for ten values of $z_m$, the maximum redshift in the quadratic interpolation. Here we use the SH0ES prior for $H_0$. The error bars are also displayed.} \label{fig: shoes_quad_zm}
\end{figure}

Using first the quadratic parametrization (\ref{xdz1}) we take several values for $z_m$ and compute the best fit values for the parameters $(\Omega_m, h,x_1,x_2)$. Let us start the analysis with the SH0ES prior for $H_0$. The results are shown in Fig.(\ref{fig: shoes_quad_zm}). Clearly, there is a trend of getting smaller values for $x_1$ and $x_2$ as we use a larger value for $z_m$.

\begin{figure}[h!]
\centering
\includegraphics[width=8cm]{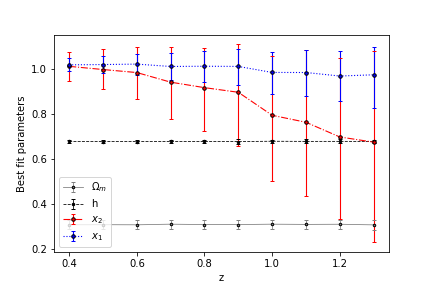}
\caption{Best fit values for $\Omega_m$, $h$, $x_1$ and $x_2$ for ten values of $z_m$, the maximum redshift in the quadratic interpolation. Here we use the Planck prior for $H_0$.} \label{fig: planck_quad_zm}
\end{figure}

In Fig.(\ref{fig: planck_quad_zm}) we display the equivalent plot for the case using the Planck prior for $H_0$. The trend is the same although the decrease is less pronounced as in the previous case. It is important to stress here that this trend continues as we use larger values of $z_m$ even reaching negative values for the best fit of $x_2$ first and then $x_1$. Also it is important to notice that the best fit values for parameters $\Omega_m$ and $h$ remain almost unchanged as we increase $z_m$. These results are again in agreement with our previous study \cite{C_rdenas_2015}, where using type Ia data and also gas mass fraction in clusters we found the data suggested that our reconstructed probe $X(z)$ decrease as the redshift increases.
%ok

Performing a series of runs of our MCMC code, we can choose an optimum $z_m$ that makes both the errors in the parameters minimal and at the same time cover enough data points to reconstruct the function $X(z)$ properly. For the quadratic parametrization with the SH0ES and Planck prior for $H_0$ we find that using $z_m=1.0$ gives us the best scope and precision in the analysis. Using this value, we obtain as the best fit parameters using SH0ES: $\Omega_m = 0.31 \pm 0.02 $, $h = 0.700 \pm 0.008 $, $x_1 = 0.87 \pm 0.09 $ and $x_2 = 0.54 \pm 0.27$. Using the Planck prior we get $\Omega_m = 0.31 \pm 0.02 $, $h = 0.679 \pm 0.008 $, $x_1 = 0.99 \pm 0.09 $ and $x_2 = 0.83 \pm 0.25$.
\begin{figure}[h!]
\centering
\includegraphics[width=9cm]{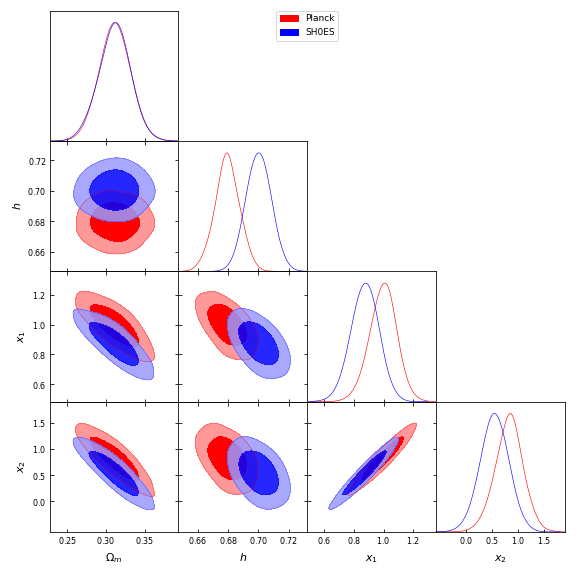}
\caption{Posteriors of the parameters using both the SH0ES prior (red contours) and the Planck prior (green contours). We notice that using the SH0ES prior the best fit values for $x_1$ and $x_2$ are appreciably displaced from 1, while using Planck these are consistent with LCDM.} \label{fig: triangle_quad}
\end{figure}
The one and two dimensional projections of the posterior probability distributions of our parameters are shown in Fig.(\ref{fig: triangle_quad}). We notice that the election of prior moves appreciably the contours on $H_0$ -- as we expected -- but also it has impact on the determination of the best fit values for the parameters $x_1$ and $x_2$. In fact, comparing the contours in the projection block $(\Omega_m, x_1)$ it is evident that the SH0ES prior points to lower than 1 values for $x_1$ (almost ruling out the LCDM value at 1 sigma), while something similar occurs with $x_2$ although the error is larger.
%ok

\begin{figure}[h!]
\centering
\includegraphics[width=9cm]{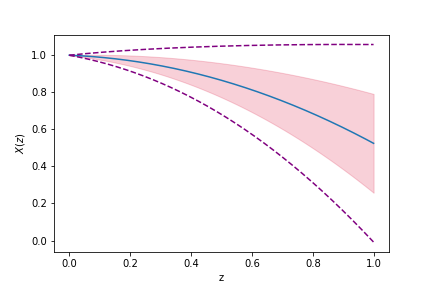}
\caption{$X(z)$ reconstruction using SH0ES prior and $z_{max}=1.0$, where the shaded region shows the 1 $\sigma$ confidence interval, and the dashed lines the 2 $\sigma$ zone.} \label{fig: wdz_shoes_zm}
\end{figure}

Using the best fit values for each case, considering the errors and the correlations among the parameters, we can reconstruct the DE density function $X(z)$. In the case of the SH0ES prior we obtain Fig. (\ref{fig: wdz_shoes_zm}). 
As we have discussed in the previous paragraph, the best fit and the reconstructed $X(z)$ is certainly not consistent with LCDM at 1 $\sigma$. Although it is close to 1 for $z<0.5$, after that the 1 $\sigma$ reconstruction falls below 1 following the trend towards a negative value for $X(z)$ for large redshifts. Noticing the vertical axes, we find also that for redshift larger than $z \simeq 1.2$ at 1 $\sigma$ the function start to take negative values for $X(z)$.
%ok

\begin{figure}[h!]
\centering
\includegraphics[width=9cm]{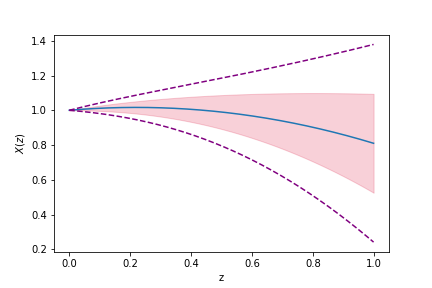}
\caption{The same as the Fig.(\ref{fig: wdz_shoes_zm}) but this time $X(z)$ is reconstructed using the Planck prior and $z_{max}=1.0$. } \label{fig: wdz_planck_zm}
\end{figure}

In the case of the Planck prior we obtain the result displayed in Fig.(\ref{fig: wdz_planck_zm}).
The graph shows that already at 1 $\sigma$ the reconstructed $X(z)$ is consistent with the LCDM value $X \simeq 1$. The reconstructed errors at 2 $\sigma$ shows no possibility to say something about DE evolution. Although the best fit reconstruction of the DE density (the thick line in Fig.(\ref{fig: wdz_planck_zm})) shows a trend similar to the SH0ES result previously mentioned, the error propagation teach us that no conclusion is possible to obtain for DE evolution.
%ok

\subsection{Cubic case}

Now we move to the cubic parametrization Eq.(\ref{xdz2}) and perform the same analysis as for the quadratic case. As before, we start studying the impact on the best fit parameters varying the maximum redshift $z_m$.
%ok

\begin{figure}[h!]
\centering
\includegraphics[width=8cm]{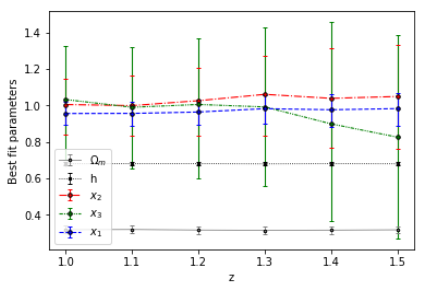}
\caption{Best fit values for $\Omega_m$, $h$, $f_1$ and $f_2$ for six values of $z_m$, the maximum redshift in the cubic interpolation. Here we use the Planck prior for $H_0$. Also the errors are displayed.} \label{fig: bestcubplanck_zm}
\end{figure}

In Fig.(\ref{fig: bestcubplanck_zm}) we show how the best fit parameters varied as we use six different values for $z_m$ using the Planck prior for $H_0$. As in the case of the quadratic parametrization, the best fit values for $\Omega_m$ and $h$ does not change appreciably as we change $z_m$. Noticing the vertical scale of Fig.(\ref{fig: bestcubplanck_zm}), we find that both $x_1$ and $x_2$ show almost no change with $z_m$, and the parameter $x_3$ although fall for $z>1.3$ the errors increase making all the fit consistent with the LCDM values.

\begin{figure}[h!]
\centering
\includegraphics[width=8cm]{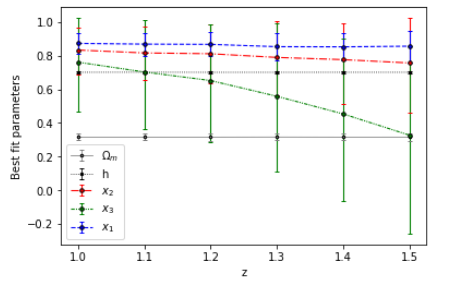}
\caption{Best fit values for $\Omega_m$, $h$, $f_1$ and $f_2$ for six values of $z_m$, the maximum redshift in the cubic interpolation. Here we use the SH0ES prior for $H_0$.} \label{fig: bestcubshoes_zm}
\end{figure}

When using the SH0ES prior, the result exhibits a decline in the cubic parameters $x_1$, $x_2$ and $x_3$ from the beginning. Noticing the vertical scale of Fig.(\ref{fig: bestcubshoes_zm}), the three parameters get values lower than 1 for all values of $z_m$. The best fit for $x_1$ shows almost no variation as $\Omega_m$ and $h$, keeping a value around $x_1 \simeq 0.88 $ well below the LCDM value according to the errors at 1 $\sigma$. For $x_2$ the best fit values show a small decrease as $z_m$ increases. Noticing the errors, the values for $x_2$ are inconsistent at 1$\sigma$ with the value 1, but the error increase for $z_m >1.3$ making it consistent with the LCDM value. The strongest effect is on $x_3$. It starts at $z_m=1$ already below the LCDM value 1, and decrease as we increase $z_m$. At 1 $\sigma$ the best fit values for $x_3$ for each $z_m$ studied are all inconsistent with the LCDM value. However this effect disappear at 2 $\sigma$.

\begin{figure}[h!]
\centering
\includegraphics[width=9cm]{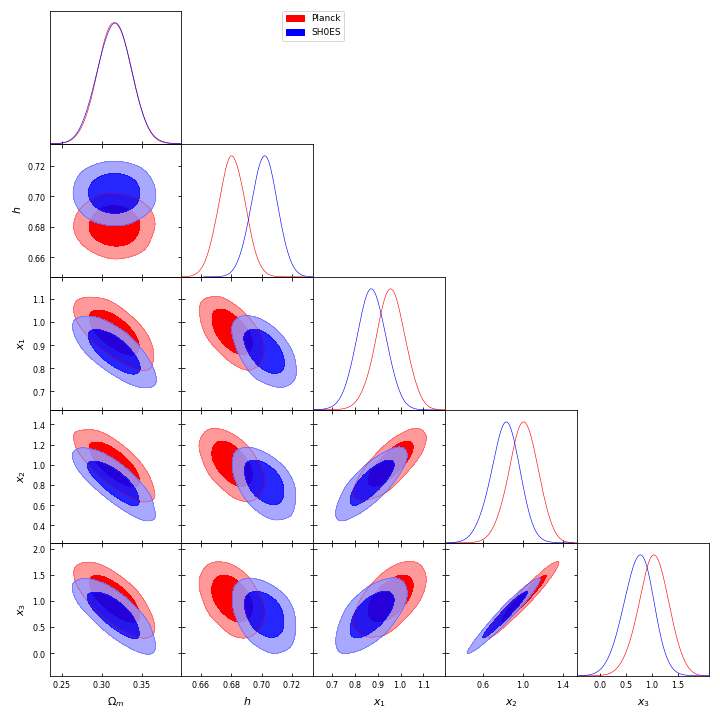}
\caption{Posteriors of the cubic parameters using both the SH0ES prior (blue contours) and the Planck prior (red contours). We notice that as in the quadratic case, the choice of prior has a significant impact on the eventual conclusion of evolution of the DE density.} \label{fig: triangle_cub}
\end{figure}

In both cases we decided to keep $z_m = 1.1$ mainly because the bulk of the data is concentrated up to that value. Performing the analysis using 5000 chains for each prior on $H_0$ we get the posteriors for 1D and 2D of the system shown in Fig.(\ref{fig: triangle_cub}).
As we can observe, we find similar results to the quadratic case. The choice of the $H_0$ prior determines if the reconstructed DE density evolves or not, at least at 1$\sigma$. The SH0ES prior, that with the larger value for $H_0$, produces the most significant evolution. In numbers the best fit values are: using the SH0ES prior we find $\Omega_m = 0.32 \pm 0.02$, $h = 0.701 \pm 0.008$, $x_1 = 0.87 \pm 0.06$, $x_2 = 0.85 \pm 0.16$, and $x_3 = 0.77 \pm 0.33$. Using the Planck prior we get $\Omega_m = 0.32 \pm 0.02$, $h = 0.680 \pm 0.008$, $x_1 = 0.96 \pm 0.07$, $x_2 = 1.04 \pm 0.17$, and $x_3 = 1.08 \pm 0.34$.
%ok

\begin{figure}[h!]
\centering
\includegraphics[width=9cm]{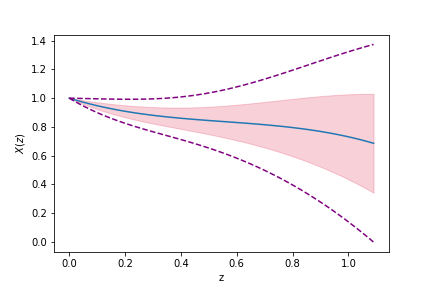}
\caption{Reconstructed energy density evolution with redshift for the cubic parametrization using SH0ES prior. The values are taken up to z=1.1 which is the maximum redshift used for this particular parametrization.} \label{fig: xdzcubshoes_zm}
\end{figure}

\begin{figure}[h!]
\centering
\includegraphics[width=9cm]{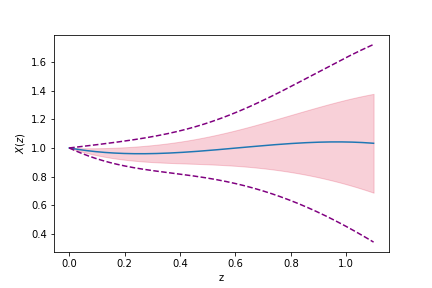}
\caption{Reconstructed energy density evolution with redshift for the cubic parametrization using Planck prior. The values are taken up to z=1.1 which is the maximum redshift used for this particular parametrization.} \label{fig: xdzcubplanck_zm}
\end{figure}

Using the best fit values of the parameters $(\Omega_m, h, x_1,x_2,x_3)$ obtained for each prior election, we can reconstruct the DE density $X(z)$ with error propagation. This is shown in Fig.(\ref{fig: xdzcubshoes_zm}) for the SH0ES prior and in Fig.(\ref{fig: xdzcubplanck_zm}) for the Planck prior. As in the quadratic case, the reconstructed $ X(z)$ up to 1 $ \sigma $ shows a clear decrease with the redshift using SH0Es prior.
Although the value of $z_m$ used in this analysis is well below the redshift at which this effect is observed, we have to stress here that in the analysis we use all the data even those for $z>z_m$, making this effect in the reconstruction a truly well supported trend. Again, as we find in the quadratic case, using the best fit values using the Planck prior for $H_0$, the reconstructed DE density show in Fig.(\ref{fig: xdzcubplanck_zm}) does not show a clear tendency away from the LCDM model.

In order to determine how good the two parametrizations methods
are in describing the data compared to $\Lambda$CDM, we implement the minimum $\chi^2$, Akaike Information Criterion (AIC) and Bayesian Information Criterion (BIC) \cite{Trotta:2008qt, Liddle:2007fy}. In Table I we summarize the results of the analysis using the definition for AIC and
BIC.

\begin{table}[h!]
\begin{tabular}{c|c|c|c|c|c|c|}
\cline{2-7}
                            & \multicolumn{3}{l|}{Planck}                       & \multicolumn{3}{l|}{SH0ES}                        \\ \hline
\multicolumn{1}{|l|}{Model} & $\chi^2_{\rm min}$ & $\Delta$AIC & $\Delta$BIC & $\chi^2_{\rm min}$ & $\Delta$AIC & $\Delta$BIC \\ \hline
\multicolumn{1}{|l|}{$\Lambda$CDM}  & 1068.23 &  0  &0& 1077.16& 0& 0 \\ \hline
\multicolumn{1}{|l|}{Q2}& 1073.82 & 0.4 & 10.4 & 1064.6 & 0.66&10.7 \\ \hline
\multicolumn{1}{|l|}{C3}& 1060.8 & -2&12.9& 1068.32 & -2.8& 12.26\\ \hline
\end{tabular}
\caption{The goodness-of-fit statistics 
$\chi^{2}_{min}$, $\Delta$AIC and $\Delta$BIC for the two priors imposed. Here Q2 stands for the quadratic parametrization and C3 for the cubic parametrization.}\label{tab:res} 
\end{table}

According to our results, the quadratic and cubic parametrizations have $\Delta$AIC $<$ 2 for both priors indicating that these models are as good as the reference $\Lambda$CDM model in describing the data. The quadratic parametrization is slightly preferred over the cubic one under AIC criteria. On the other hand, for the BIC estimator which is well known that penalizes free extra parameters more than AIC, we obtain $\Delta$BIC $>$ 10 for both parametrizations and priors showing that there is strong evidence against the parametrization methods studied in this work.
%ok
The fact that AIC criteria differs from the BIC one is not a surprise. For example, the reconstruction made in \cite{mehrabi_2018} using the Padè approximation have exactly the same behavior. In fact, given the penalty shown by the statistical tests, there is no gain in including more complex models in the parametrization. This is clearly shown also in \cite{mehrabi_2018} where there is a strong evidence against models with more than two free parameters.

%%%%%%%%%%%%%%%%%%%%%%%%%%%%%%%%%%%%%%%%%%%%%%%%%%%%%%%%%%
\section{\label{sec: theoryideas} Theoretical ideas}

Because the results we have obtained in the previous sections indicate a mild trend for $X(z)$ decreasing as redshift increase, in this section we discuss some theoretical models that may produce such a behavior.
%ok

\begin{figure}[h]
    \includegraphics[width=8cm]{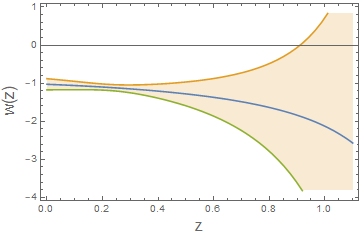}
    \includegraphics[width=8cm]{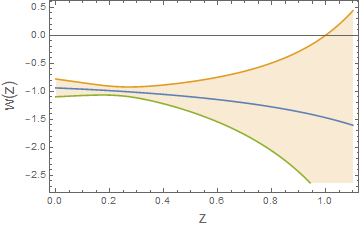}
\caption{The reconstructed EoS parameter $w(z)$ for quadratic parametrization. The upper panel correspond to the SH0ES prior and the bottom to the Planck one. }\label{fig: wdzrecquad}
\end{figure}

Let us start characterizing this behavior. Using the relation (\ref{wdz}) we can reconstruct the EoS parameter as a function of redshift by
\begin{equation}\label{wdxz}
w(z) = -1 + \frac{1+z}{3}\frac{X'(z)}{X(z)}.
\end{equation}
Having found the best fit values for $X(z)$, we can use it to compute any derivative of $X(z)$ and plot the result with the proper error propagation. Of course this can be done for each parametrization and also for each prior on $H_0$. The results for the quadratic case are shown in Fig. (\ref{fig: wdzrecquad}).
Both graphs shows similar behavior. Although the best fit evolution is close to $w=-1$ today, it slowly falls below that line, into the phantom regime, as redshift increases. Notice also the vertical scale, showing that the reconstruction for the SH0ES case shows a larger spread compared to that of Planck. However, up to one sigma, both results are in agreement with the cosmological constant result.
%ok

\begin{figure}[h]
    \includegraphics[width=8cm]{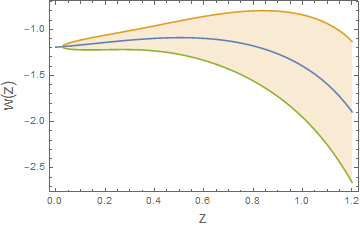}
    \includegraphics[width=8cm]{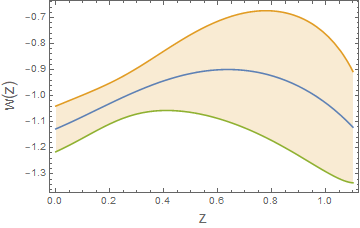}
\caption{The reconstructed $w(z)$ for the cubic parametrization. The upper panel correspond to the SH0ES prior and the bottom to Planck. }\label{fig: wdzreccubic}
\end{figure}

In contrast, for the case of the cubic parametrization, the evolution found differs notably respect to the cosmological constant. The results in Fig. (\ref{fig: wdzreccubic}), show that as redshift increases the reconstructed $w(z)$ varies appreciably. The best fit evolution, in both cases, are of phantom type today, then evolves, as redshift increases,  to less negative values, reaches certain maximum and then decreases towards more negative values. 
Notice also the vertical scale, again the SH0ES case shows a large spread in the reconstructed values for $w(z)$ compared to Planck. Even up to one sigma the evolution of $w(z)$ is noticeably, in both cases showing a phantom evolution for $z$ larger than 1.
%ok

In the same way we have reconstructed $w(z)$, we can also do this with other physical quantities as the deceleration parameter $q(z)$ for example, or even the interaction function $Q(z)$. In fact, we can use the methods described in \cite{Hwei}, \cite{RRos} where we can built it from $f(a)=\rho_X/\rho_m$ through
\begin{equation}
Q(a) = -H\rho_m \Omega_X \left(a\frac{f'}{f} + 3w_x \right).
\end{equation}
where a prime means derivative with respect to the scale factor $a$, and $\Omega_X = f/(1+f) = 1 - \Omega_m$.
%ok

Both the decreasing trend of $X(z)$ as redshift increases and that of $w(z)$ shown above are features observed in interacting models of dark matter and dark energy. In the next sub section we study a sample example of this type.
%ok

\subsection{Dark interactions}

\begin{figure}[h]
\centering
\includegraphics[width=8cm]{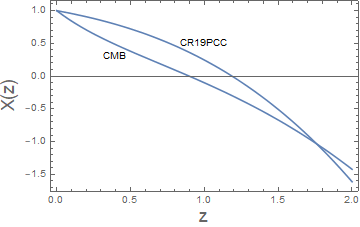}
\caption{We display (\ref{xdzteo}) for two of the best fits informed in Ref.\cite{Yang:2020uga}. We have kept the label of the table in the reference. See the text for details. Notice the trend of the DE density with redshift, reaching even negative values for $z>1.$ }\label{fig: teoria}
\end{figure}

Let us discuss here as an example, a model with explicit interaction between DM and DE. The equations governing the system take the form
\begin{eqnarray}
3H^2 = \rho_m + \rho_{de}, \\
\dot{\rho}_m + 3 H \rho_m = -Q, \\
\dot{\rho}_{de} + 3 H (\rho_{de} + p_{de}) = Q,
\end{eqnarray}
Assuming an interaction function of the type $Q= 3H\xi \rho_{de}$ -- which is the preferred function based on observations according to \cite{Yang:2020uga} -- we obtain 
\begin{equation}
    \rho_{de}(a) = \rho_{de}(1) a^{-3(1+w_{de}-\xi)} ,
\end{equation}
and using it in the equation for $\rho_m$ we obtain
\begin{equation}
    \rho_m (a) = \rho_m(1)a^{-3} + \frac{\xi \rho_{de}(1)a^{-3}}{\xi - w_{de}}\left[ 1-a^{-3(w_{de}-\xi)}\right].
\end{equation}
Once written in the Hubble function we can identify the effective $X(a)$ as the term that accompanies the term of pure non-relativistic matter that goes as $a^{-3}$. Explicitly it takes the form
\begin{equation}
X(a) = \frac{\xi a^{-3} - w_{de} a^{-3(1+w_{de}-\xi)}}{\xi -w_{de}} .\label{xdzteo}
\end{equation}
Although the first term increases as redshift increases (because $a = (1+z)^{-1}$), the best fit values quoted in \cite{Yang:2020uga} for $\xi$ are all negative, then the first term really decreases towards negative values, while the second term depends on the value of the combination $\Delta = 1+w_{de}-\xi$. Typically the interaction parameter $\xi$ is small and negative, for phantom like prior. Having $w_{de} < -1$, we get a small positive $\Delta $. In this case the second term is essentially a constant that level up the behavior imposed by the first term, producing a $X(z)$ that decrease as redshift increases.
We take two sets of best fit values from Table II \cite{Yang:2020uga}, where typically $\xi \simeq -0.1$, and $w_{de} \simeq -1.1$ and plot the results of inserting them in (\ref{xdzteo}) Fig.(\ref{fig: teoria}). %
\\
\\

\section{\label{sec: conclusions} Conclusions}

We have reconstructed the DE density $X(z)$ defined in (\ref{edz}) from type Ia supernova and $H(z)$ measurements using a quadratic and cubic parametrization. In the analysis for both cases we have used two priors for $H_0$ separately, the one from Planck \cite{Aghanim:2018eyx} and from SH0ES \cite{Riess:2019cxk}. We found that the best fit values points to a evolution of $X(z)$ with redshift – decreasing as redshift increase – keeping the trend found first in \cite{C_rdenas_2015}. Extending the analysis up to 1$\sigma$, the best fits are still consistent with evolution in the case of using the SH0ES prior for the quadratic parametrization, but disappears from the fit using the Planck prior in both parametrizations. Extending the analysis even more, up to 2$\sigma$, the evidence for evolution of $X(z)$ dilute in all cases. 
%ok

Although the statistical significance of this potential evolution is still low, we take this evidence seriously as a clue to develop an eventual model. We also check the statistical performance of the parametrization models studied in this work by computing the $\chi^2_{\rm min}$, AIC and BIC estimators. There is no important difference in the $\chi^2_{\rm min}$ obtained for the different models, and the AIC criteria shows that it is not possible to distinguish between $\Lambda$CDM and the parametrization models, hence the possibility of a dynamical dark energy under these parametrization forms should be explored further. However, BIC estimator penalizes model complexity more heavily and shows strong evidence against models Q2 and C3, so under this criteria the model $\Lambda$CDM is preferred in describing the data.

We study as an example the case of an interaction model between dark components. Using the best estimate for the interaction parameters we find that assuming a phantomlike EoS for DE, as was done in Ref.\cite{Yang:2020uga}, we obtain exactly the same trend we have found in this work, a DE density decreasing as redshift increases.
%ok

\begin{acknowledgments}
DG acknowledges support from project ANID-PFCHA/Doctorado Nacional/2019-21191886. The authors would like to thank the reviewers for their comments towards improving our manuscript.
\end{acknowledgments}

\bibliography{deevolv}% Produces the bibliography via BibTeX.

%apsrev4-2.bst 2019-01-14 (MD) hand-edited version of apsrev4-1.bst
%Control: key (0)
%Control: author (8) initials jnrlst
%Control: editor formatted (1) identically to author
%Control: production of article title (0) allowed
%Control: page (0) single
%Control: year (1) truncated
%Control: production of eprint (0) enabled
\providecommand{\noopsort}[1]{}\providecommand{\singleletter}[1]{#1}%
\begin{thebibliography}{43}%
\makeatletter
\providecommand \@ifxundefined [1]{%
 \@ifx{#1\undefined}
}%
\providecommand \@ifnum [1]{%
 \ifnum #1\expandafter \@firstoftwo
 \else \expandafter \@secondoftwo
 \fi
}%
\providecommand \@ifx [1]{%
 \ifx #1\expandafter \@firstoftwo
 \else \expandafter \@secondoftwo
 \fi
}%
\providecommand \natexlab [1]{#1}%
\providecommand \enquote  [1]{``#1''}%
\providecommand \bibnamefont  [1]{#1}%
\providecommand \bibfnamefont [1]{#1}%
\providecommand \citenamefont [1]{#1}%
\providecommand \href@noop [0]{\@secondoftwo}%
\providecommand \href [0]{\begingroup \@sanitize@url \@href}%
\providecommand \@href[1]{\@@startlink{#1}\@@href}%
\providecommand \@@href[1]{\endgroup#1\@@endlink}%
\providecommand \@sanitize@url [0]{\catcode `\\12\catcode `\$12\catcode
  `\&12\catcode `\#12\catcode `\^12\catcode `\_12\catcode `\%12\relax}%
\providecommand \@@startlink[1]{}%
\providecommand \@@endlink[0]{}%
\providecommand \url  [0]{\begingroup\@sanitize@url \@url }%
\providecommand \@url [1]{\endgroup\@href {#1}{\urlprefix }}%
\providecommand \urlprefix  [0]{URL }%
\providecommand \Eprint [0]{\href }%
\providecommand \doibase [0]{https://doi.org/}%
\providecommand \selectlanguage [0]{\@gobble}%
\providecommand \bibinfo  [0]{\@secondoftwo}%
\providecommand \bibfield  [0]{\@secondoftwo}%
\providecommand \translation [1]{[#1]}%
\providecommand \BibitemOpen [0]{}%
\providecommand \bibitemStop [0]{}%
\providecommand \bibitemNoStop [0]{.\EOS\space}%
\providecommand \EOS [0]{\spacefactor3000\relax}%
\providecommand \BibitemShut  [1]{\csname bibitem#1\endcsname}%
\let\auto@bib@innerbib\@empty
%</preamble>
\bibitem [{\citenamefont {Riess}\ \emph {et~al.}(1998)\citenamefont {Riess}
  \emph {et~al.}}]{Riess:1998cb}%
  \BibitemOpen
  \bibfield  {author} {\bibinfo {author} {\bibfnamefont {A.~G.}\ \bibnamefont
  {Riess}} \emph {et~al.} (\bibinfo {collaboration} {Supernova Search Team}),\
  }\href {https://doi.org/10.1086/300499} {\bibfield  {journal} {\bibinfo
  {journal} {Astron. J.}\ }\textbf {\bibinfo {volume} {116}},\ \bibinfo {pages}
  {1009} (\bibinfo {year} {1998})},\ \Eprint
  {https://arxiv.org/abs/astro-ph/9805201} {arXiv:astro-ph/9805201}
  \BibitemShut {NoStop}%
\bibitem [{\citenamefont {Perlmutter}\ \emph {et~al.}(1999)\citenamefont
  {Perlmutter}, \citenamefont {Aldering}, \citenamefont {Goldhaber},
  \citenamefont {Knop}, \citenamefont {Nugent}, \citenamefont {Castro},
  \citenamefont {Deustua}, \citenamefont {Fabbro}, \citenamefont {Goobar},
  \citenamefont {Groom},\ and\ \citenamefont {et~al.}}]{Perlmutter_1999}%
  \BibitemOpen
  \bibfield  {author} {\bibinfo {author} {\bibfnamefont {S.}~\bibnamefont
  {Perlmutter}}, \bibinfo {author} {\bibfnamefont {G.}~\bibnamefont
  {Aldering}}, \bibinfo {author} {\bibfnamefont {G.}~\bibnamefont {Goldhaber}},
  \bibinfo {author} {\bibfnamefont {R.~A.}\ \bibnamefont {Knop}}, \bibinfo
  {author} {\bibfnamefont {P.}~\bibnamefont {Nugent}}, \bibinfo {author}
  {\bibfnamefont {P.~G.}\ \bibnamefont {Castro}}, \bibinfo {author}
  {\bibfnamefont {S.}~\bibnamefont {Deustua}}, \bibinfo {author} {\bibfnamefont
  {S.}~\bibnamefont {Fabbro}}, \bibinfo {author} {\bibfnamefont
  {A.}~\bibnamefont {Goobar}}, \bibinfo {author} {\bibfnamefont {D.~E.}\
  \bibnamefont {Groom}},\ and\ \bibinfo {author} {\bibnamefont {et~al.}},\
  }\href {https://doi.org/10.1086/307221} {\bibfield  {journal} {\bibinfo
  {journal} {The Astrophysical Journal}\ }\textbf {\bibinfo {volume} {517}},\
  \bibinfo {pages} {565–586} (\bibinfo {year} {1999})}\BibitemShut {NoStop}%
\bibitem [{\citenamefont {Frieman}\ \emph {et~al.}(2008)\citenamefont
  {Frieman}, \citenamefont {Turner},\ and\ \citenamefont
  {Huterer}}]{Frieman:2008sn}%
  \BibitemOpen
  \bibfield  {author} {\bibinfo {author} {\bibfnamefont {J.}~\bibnamefont
  {Frieman}}, \bibinfo {author} {\bibfnamefont {M.}~\bibnamefont {Turner}},\
  and\ \bibinfo {author} {\bibfnamefont {D.}~\bibnamefont {Huterer}},\ }\href
  {https://doi.org/10.1146/annurev.astro.46.060407.145243} {\bibfield
  {journal} {\bibinfo  {journal} {Ann. Rev. Astron. Astrophys.}\ }\textbf
  {\bibinfo {volume} {46}},\ \bibinfo {pages} {385} (\bibinfo {year} {2008})},\
  \Eprint {https://arxiv.org/abs/0803.0982} {arXiv:0803.0982 [astro-ph]}
  \BibitemShut {NoStop}%
\bibitem [{\citenamefont {Huterer}\ and\ \citenamefont
  {Shafer}(2017)}]{Huterer_2017}%
  \BibitemOpen
  \bibfield  {author} {\bibinfo {author} {\bibfnamefont {D.}~\bibnamefont
  {Huterer}}\ and\ \bibinfo {author} {\bibfnamefont {D.~L.}\ \bibnamefont
  {Shafer}},\ }\href {https://doi.org/10.1088/1361-6633/aa997e} {\bibfield
  {journal} {\bibinfo  {journal} {Reports on Progress in Physics}\ }\textbf
  {\bibinfo {volume} {81}},\ \bibinfo {pages} {016901} (\bibinfo {year}
  {2017})}\BibitemShut {NoStop}%
\bibitem [{\citenamefont {Chevallier}\ and\ \citenamefont
  {Polarski}(2001)}]{Chevallier:2000qy}%
  \BibitemOpen
  \bibfield  {author} {\bibinfo {author} {\bibfnamefont {M.}~\bibnamefont
  {Chevallier}}\ and\ \bibinfo {author} {\bibfnamefont {D.}~\bibnamefont
  {Polarski}},\ }\href {https://doi.org/10.1142/S0218271801000822} {\bibfield
  {journal} {\bibinfo  {journal} {Int. J. Mod. Phys. D}\ }\textbf {\bibinfo
  {volume} {10}},\ \bibinfo {pages} {213} (\bibinfo {year} {2001})},\ \Eprint
  {https://arxiv.org/abs/gr-qc/0009008} {arXiv:gr-qc/0009008} \BibitemShut
  {NoStop}%
\bibitem [{\citenamefont {Linder}(2003)}]{Linder:2002et}%
  \BibitemOpen
  \bibfield  {author} {\bibinfo {author} {\bibfnamefont {E.~V.}\ \bibnamefont
  {Linder}},\ }\href {https://doi.org/10.1103/PhysRevLett.90.091301} {\bibfield
   {journal} {\bibinfo  {journal} {Phys. Rev. Lett.}\ }\textbf {\bibinfo
  {volume} {90}},\ \bibinfo {pages} {091301} (\bibinfo {year} {2003})},\
  \Eprint {https://arxiv.org/abs/astro-ph/0208512} {arXiv:astro-ph/0208512}
  \BibitemShut {NoStop}%
\bibitem [{\citenamefont {Maor}\ \emph {et~al.}(2001)\citenamefont {Maor},
  \citenamefont {Brustein},\ and\ \citenamefont {Steinhardt}}]{Maor_2001}%
  \BibitemOpen
  \bibfield  {author} {\bibinfo {author} {\bibfnamefont {I.}~\bibnamefont
  {Maor}}, \bibinfo {author} {\bibfnamefont {R.}~\bibnamefont {Brustein}},\
  and\ \bibinfo {author} {\bibfnamefont {P.~J.}\ \bibnamefont {Steinhardt}},\
  }\bibfield  {title} {\bibinfo {title} {Limitations in using luminosity
  distance to determine the equation of state of the universe},\ }\href
  {https://doi.org/10.1103/physrevlett.86.6} {\bibfield  {journal} {\bibinfo
  {journal} {Physical Review Letters}\ }\textbf {\bibinfo {volume} {86}},\
  \bibinfo {pages} {6–9} (\bibinfo {year} {2001})}\BibitemShut {NoStop}%
\bibitem [{\citenamefont {Tegmark}(2002)}]{Tegmark_2002}%
  \BibitemOpen
  \bibfield  {author} {\bibinfo {author} {\bibfnamefont {M.}~\bibnamefont
  {Tegmark}},\ }\bibfield  {title} {\bibinfo {title} {Measuring the metric: A
  parametrized post-friedmannian approach to the cosmic dark energy problem},\
  }\bibfield  {journal} {\bibinfo  {journal} {Physical Review D}\ }\textbf
  {\bibinfo {volume} {66}},\ \href {https://doi.org/10.1103/physrevd.66.103507}
  {10.1103/physrevd.66.103507} (\bibinfo {year} {2002})\BibitemShut {NoStop}%
\bibitem [{\citenamefont {Huterer}\ and\ \citenamefont
  {Starkman}(2003)}]{Huterer:2002hy}%
  \BibitemOpen
  \bibfield  {author} {\bibinfo {author} {\bibfnamefont {D.}~\bibnamefont
  {Huterer}}\ and\ \bibinfo {author} {\bibfnamefont {G.}~\bibnamefont
  {Starkman}},\ }\href {https://doi.org/10.1103/PhysRevLett.90.031301}
  {\bibfield  {journal} {\bibinfo  {journal} {Phys. Rev. Lett.}\ }\textbf
  {\bibinfo {volume} {90}},\ \bibinfo {pages} {031301} (\bibinfo {year}
  {2003})},\ \Eprint {https://arxiv.org/abs/astro-ph/0207517}
  {arXiv:astro-ph/0207517} \BibitemShut {NoStop}%
\bibitem [{\citenamefont {Wang}\ and\ \citenamefont
  {Lovelace}(2002)}]{Wang_2002}%
  \BibitemOpen
  \bibfield  {author} {\bibinfo {author} {\bibfnamefont {Y.}~\bibnamefont
  {Wang}}\ and\ \bibinfo {author} {\bibfnamefont {G.}~\bibnamefont
  {Lovelace}},\ }\bibfield  {title} {\bibinfo {title} {Unbiased estimate of
  dark energy density from type i[{CLC}]a[/{CLC}] supernova data},\ }\href
  {https://doi.org/10.1086/338142} {\bibfield  {journal} {\bibinfo  {journal}
  {The Astrophysical Journal}\ }\textbf {\bibinfo {volume} {562}},\ \bibinfo
  {pages} {L115} (\bibinfo {year} {2002})}\BibitemShut {NoStop}%
\bibitem [{\citenamefont {Wang}\ and\ \citenamefont
  {Garnavich}(2001)}]{Wang:2001ht}%
  \BibitemOpen
  \bibfield  {author} {\bibinfo {author} {\bibfnamefont {Y.}~\bibnamefont
  {Wang}}\ and\ \bibinfo {author} {\bibfnamefont {P.~M.}\ \bibnamefont
  {Garnavich}},\ }\href {https://doi.org/10.1086/320552} {\bibfield  {journal}
  {\bibinfo  {journal} {Astrophys. J.}\ }\textbf {\bibinfo {volume} {552}},\
  \bibinfo {pages} {445} (\bibinfo {year} {2001})},\ \Eprint
  {https://arxiv.org/abs/astro-ph/0101040} {arXiv:astro-ph/0101040}
  \BibitemShut {NoStop}%
\bibitem [{\citenamefont {Wang}\ and\ \citenamefont
  {Mukherjee}(2004)}]{Wang:2003gz}%
  \BibitemOpen
  \bibfield  {author} {\bibinfo {author} {\bibfnamefont {Y.}~\bibnamefont
  {Wang}}\ and\ \bibinfo {author} {\bibfnamefont {P.}~\bibnamefont
  {Mukherjee}},\ }\href {https://doi.org/10.1086/383196} {\bibfield  {journal}
  {\bibinfo  {journal} {Astrophys. J.}\ }\textbf {\bibinfo {volume} {606}},\
  \bibinfo {pages} {654} (\bibinfo {year} {2004})},\ \Eprint
  {https://arxiv.org/abs/astro-ph/0312192} {arXiv:astro-ph/0312192}
  \BibitemShut {NoStop}%
\bibitem [{\citenamefont {Wang}\ and\ \citenamefont
  {Freese}(2006)}]{Wang:2004ru}%
  \BibitemOpen
  \bibfield  {author} {\bibinfo {author} {\bibfnamefont {Y.}~\bibnamefont
  {Wang}}\ and\ \bibinfo {author} {\bibfnamefont {K.}~\bibnamefont {Freese}},\
  }\href {https://doi.org/10.1016/j.physletb.2005.10.083} {\bibfield  {journal}
  {\bibinfo  {journal} {Phys. Lett. B}\ }\textbf {\bibinfo {volume} {632}},\
  \bibinfo {pages} {449} (\bibinfo {year} {2006})},\ \Eprint
  {https://arxiv.org/abs/astro-ph/0402208} {arXiv:astro-ph/0402208}
  \BibitemShut {NoStop}%
\bibitem [{\citenamefont {mehrabi}\ and\ \citenamefont
  {Basilakos}(2018)}]{mehrabi_2018}%
  \BibitemOpen
  \bibfield  {author} {\bibinfo {author} {\bibfnamefont {A.}~\bibnamefont
  {mehrabi}}\ and\ \bibinfo {author} {\bibfnamefont {S.}~\bibnamefont
  {Basilakos}},\ }\bibfield  {title} {\bibinfo {title} {Dark energy
  reconstruction based on the padé approximation; an expansion around the
  $\lambda$cdm},\ }\bibfield  {journal} {\bibinfo  {journal} {The European
  Physical Journal C}\ }\textbf {\bibinfo {volume} {78}},\ \href
  {https://doi.org/10.1140/epjc/s10052-018-6368-x}
  {10.1140/epjc/s10052-018-6368-x} (\bibinfo {year} {2018})\BibitemShut
  {NoStop}%
\bibitem [{\citenamefont {Holsclaw}\ \emph
  {et~al.}(2010{\natexlab{a}})\citenamefont {Holsclaw}, \citenamefont {Alam},
  \citenamefont {Sanso}, \citenamefont {Lee}, \citenamefont {Heitmann},
  \citenamefont {Habib},\ and\ \citenamefont {Higdon}}]{Holsclaw:2010nb}%
  \BibitemOpen
  \bibfield  {author} {\bibinfo {author} {\bibfnamefont {T.}~\bibnamefont
  {Holsclaw}}, \bibinfo {author} {\bibfnamefont {U.}~\bibnamefont {Alam}},
  \bibinfo {author} {\bibfnamefont {B.}~\bibnamefont {Sanso}}, \bibinfo
  {author} {\bibfnamefont {H.}~\bibnamefont {Lee}}, \bibinfo {author}
  {\bibfnamefont {K.}~\bibnamefont {Heitmann}}, \bibinfo {author}
  {\bibfnamefont {S.}~\bibnamefont {Habib}},\ and\ \bibinfo {author}
  {\bibfnamefont {D.}~\bibnamefont {Higdon}},\ }\bibfield  {title} {\bibinfo
  {title} {{Nonparametric Reconstruction of the Dark Energy Equation of
  State}},\ }\href {https://doi.org/10.1103/PhysRevD.82.103502} {\bibfield
  {journal} {\bibinfo  {journal} {Phys. Rev. D}\ }\textbf {\bibinfo {volume}
  {82}},\ \bibinfo {pages} {103502} (\bibinfo {year} {2010}{\natexlab{a}})},\
  \Eprint {https://arxiv.org/abs/1009.5443} {arXiv:1009.5443 [astro-ph.CO]}
  \BibitemShut {NoStop}%
\bibitem [{\citenamefont {Holsclaw}\ \emph
  {et~al.}(2010{\natexlab{b}})\citenamefont {Holsclaw}, \citenamefont {Alam},
  \citenamefont {Sanso}, \citenamefont {Lee}, \citenamefont {Heitmann},
  \citenamefont {Habib},\ and\ \citenamefont {Higdon}}]{Holsclaw:2010sk}%
  \BibitemOpen
  \bibfield  {author} {\bibinfo {author} {\bibfnamefont {T.}~\bibnamefont
  {Holsclaw}}, \bibinfo {author} {\bibfnamefont {U.}~\bibnamefont {Alam}},
  \bibinfo {author} {\bibfnamefont {B.}~\bibnamefont {Sanso}}, \bibinfo
  {author} {\bibfnamefont {H.}~\bibnamefont {Lee}}, \bibinfo {author}
  {\bibfnamefont {K.}~\bibnamefont {Heitmann}}, \bibinfo {author}
  {\bibfnamefont {S.}~\bibnamefont {Habib}},\ and\ \bibinfo {author}
  {\bibfnamefont {D.}~\bibnamefont {Higdon}},\ }\href
  {https://doi.org/10.1103/PhysRevLett.105.241302} {\bibfield  {journal}
  {\bibinfo  {journal} {Phys. Rev. Lett.}\ }\textbf {\bibinfo {volume} {105}},\
  \bibinfo {pages} {241302} (\bibinfo {year} {2010}{\natexlab{b}})},\ \Eprint
  {https://arxiv.org/abs/1011.3079} {arXiv:1011.3079 [astro-ph.CO]}
  \BibitemShut {NoStop}%
\bibitem [{\citenamefont {Crittenden}\ \emph {et~al.}(2012)\citenamefont
  {Crittenden}, \citenamefont {Zhao}, \citenamefont {Pogosian}, \citenamefont
  {Samushia},\ and\ \citenamefont {Zhang}}]{Crittenden:2011aa}%
  \BibitemOpen
  \bibfield  {author} {\bibinfo {author} {\bibfnamefont {R.~G.}\ \bibnamefont
  {Crittenden}}, \bibinfo {author} {\bibfnamefont {G.-B.}\ \bibnamefont
  {Zhao}}, \bibinfo {author} {\bibfnamefont {L.}~\bibnamefont {Pogosian}},
  \bibinfo {author} {\bibfnamefont {L.}~\bibnamefont {Samushia}},\ and\
  \bibinfo {author} {\bibfnamefont {X.}~\bibnamefont {Zhang}},\ }\href
  {https://doi.org/10.1088/1475-7516/2012/02/048} {\bibfield  {journal}
  {\bibinfo  {journal} {JCAP}\ }\textbf {\bibinfo {volume} {02}},\ \bibinfo
  {pages} {048}},\ \Eprint {https://arxiv.org/abs/1112.1693} {arXiv:1112.1693
  [astro-ph.CO]} \BibitemShut {NoStop}%
\bibitem [{\citenamefont {Zhai}\ and\ \citenamefont
  {Wang}(2019)}]{Zhai:2018vmm}%
  \BibitemOpen
  \bibfield  {author} {\bibinfo {author} {\bibfnamefont {Z.}~\bibnamefont
  {Zhai}}\ and\ \bibinfo {author} {\bibfnamefont {Y.}~\bibnamefont {Wang}},\
  }\bibfield  {title} {\bibinfo {title} {{Robust and model-independent
  cosmological constraints from distance measurements}},\ }\href
  {https://doi.org/10.1088/1475-7516/2019/07/005} {\bibfield  {journal}
  {\bibinfo  {journal} {JCAP}\ }\textbf {\bibinfo {volume} {07}},\ \bibinfo
  {pages} {005}},\ \Eprint {https://arxiv.org/abs/1811.07425} {arXiv:1811.07425
  [astro-ph.CO]} \BibitemShut {NoStop}%
\bibitem [{\citenamefont {Heisenberg}\ \emph {et~al.}(2021)\citenamefont
  {Heisenberg}, \citenamefont {Bartelmann}, \citenamefont {Brandenberger},\
  and\ \citenamefont {Refregier}}]{Heisenberg:2020ywd}%
  \BibitemOpen
  \bibfield  {author} {\bibinfo {author} {\bibfnamefont {L.}~\bibnamefont
  {Heisenberg}}, \bibinfo {author} {\bibfnamefont {M.}~\bibnamefont
  {Bartelmann}}, \bibinfo {author} {\bibfnamefont {R.}~\bibnamefont
  {Brandenberger}},\ and\ \bibinfo {author} {\bibfnamefont {A.}~\bibnamefont
  {Refregier}},\ }\bibfield  {title} {\bibinfo {title} {{Model independent
  analysis of supernova data, dark energy, trans-Planckian censorship and the
  swampland}},\ }\href {https://doi.org/10.1016/j.physletb.2020.135990}
  {\bibfield  {journal} {\bibinfo  {journal} {Phys. Lett. B}\ }\textbf
  {\bibinfo {volume} {812}},\ \bibinfo {pages} {135990} (\bibinfo {year}
  {2021})},\ \Eprint {https://arxiv.org/abs/2003.13283} {arXiv:2003.13283
  [hep-th]} \BibitemShut {NoStop}%
\bibitem [{\citenamefont {Cárdenas}(2015)}]{C_rdenas_2015}%
  \BibitemOpen
  \bibfield  {author} {\bibinfo {author} {\bibfnamefont {V.~H.}\ \bibnamefont
  {Cárdenas}},\ }\href {https://doi.org/10.1016/j.physletb.2015.08.064}
  {\bibfield  {journal} {\bibinfo  {journal} {Physics Letters B}\ }\textbf
  {\bibinfo {volume} {750}},\ \bibinfo {pages} {128–134} (\bibinfo {year}
  {2015})}\BibitemShut {NoStop}%
\bibitem [{\citenamefont {Scolnic}\ \emph {et~al.}(2018)\citenamefont
  {Scolnic}, \citenamefont {Jones}, \citenamefont {Rest}, \citenamefont {Pan},
  \citenamefont {Chornock}, \citenamefont {Foley}, \citenamefont {Huber},
  \citenamefont {Kessler}, \citenamefont {Narayan}, \citenamefont {Riess},\
  and\ \citenamefont {et~al.}}]{Scolnic_2018}%
  \BibitemOpen
  \bibfield  {author} {\bibinfo {author} {\bibfnamefont {D.~M.}\ \bibnamefont
  {Scolnic}}, \bibinfo {author} {\bibfnamefont {D.~O.}\ \bibnamefont {Jones}},
  \bibinfo {author} {\bibfnamefont {A.}~\bibnamefont {Rest}}, \bibinfo {author}
  {\bibfnamefont {Y.~C.}\ \bibnamefont {Pan}}, \bibinfo {author} {\bibfnamefont
  {R.}~\bibnamefont {Chornock}}, \bibinfo {author} {\bibfnamefont {R.~J.}\
  \bibnamefont {Foley}}, \bibinfo {author} {\bibfnamefont {M.~E.}\ \bibnamefont
  {Huber}}, \bibinfo {author} {\bibfnamefont {R.}~\bibnamefont {Kessler}},
  \bibinfo {author} {\bibfnamefont {G.}~\bibnamefont {Narayan}}, \bibinfo
  {author} {\bibfnamefont {A.~G.}\ \bibnamefont {Riess}},\ and\ \bibinfo
  {author} {\bibnamefont {et~al.}},\ }\href
  {https://doi.org/10.3847/1538-4357/aab9bb} {\bibfield  {journal} {\bibinfo
  {journal} {The Astrophysical Journal}\ }\textbf {\bibinfo {volume} {859}},\
  \bibinfo {pages} {101} (\bibinfo {year} {2018})}\BibitemShut {NoStop}%
\bibitem [{\citenamefont {Zhao}\ \emph {et~al.}(2012)\citenamefont {Zhao},
  \citenamefont {Crittenden}, \citenamefont {Pogosian},\ and\ \citenamefont
  {Zhang}}]{Zhao:2012aw}%
  \BibitemOpen
  \bibfield  {author} {\bibinfo {author} {\bibfnamefont {G.-B.}\ \bibnamefont
  {Zhao}}, \bibinfo {author} {\bibfnamefont {R.~G.}\ \bibnamefont
  {Crittenden}}, \bibinfo {author} {\bibfnamefont {L.}~\bibnamefont
  {Pogosian}},\ and\ \bibinfo {author} {\bibfnamefont {X.}~\bibnamefont
  {Zhang}},\ }\href {https://doi.org/10.1103/PhysRevLett.109.171301} {\bibfield
   {journal} {\bibinfo  {journal} {Phys. Rev. Lett.}\ }\textbf {\bibinfo
  {volume} {109}},\ \bibinfo {pages} {171301} (\bibinfo {year} {2012})},\
  \Eprint {https://arxiv.org/abs/1207.3804} {arXiv:1207.3804 [astro-ph.CO]}
  \BibitemShut {NoStop}%
\bibitem [{\citenamefont {Conley}\ \emph {et~al.}(2010)\citenamefont {Conley},
  \citenamefont {Guy}, \citenamefont {Sullivan}, \citenamefont {Regnault},
  \citenamefont {Astier}, \citenamefont {Balland}, \citenamefont {Basa},
  \citenamefont {Carlberg}, \citenamefont {Fouchez}, \citenamefont {Hardin},\
  and\ \citenamefont {et~al.}}]{Conley_2010}%
  \BibitemOpen
  \bibfield  {author} {\bibinfo {author} {\bibfnamefont {A.}~\bibnamefont
  {Conley}}, \bibinfo {author} {\bibfnamefont {J.}~\bibnamefont {Guy}},
  \bibinfo {author} {\bibfnamefont {M.}~\bibnamefont {Sullivan}}, \bibinfo
  {author} {\bibfnamefont {N.}~\bibnamefont {Regnault}}, \bibinfo {author}
  {\bibfnamefont {P.}~\bibnamefont {Astier}}, \bibinfo {author} {\bibfnamefont
  {C.}~\bibnamefont {Balland}}, \bibinfo {author} {\bibfnamefont
  {S.}~\bibnamefont {Basa}}, \bibinfo {author} {\bibfnamefont {R.~G.}\
  \bibnamefont {Carlberg}}, \bibinfo {author} {\bibfnamefont {D.}~\bibnamefont
  {Fouchez}}, \bibinfo {author} {\bibfnamefont {D.}~\bibnamefont {Hardin}},\
  and\ \bibinfo {author} {\bibnamefont {et~al.}},\ }\href
  {https://doi.org/10.1088/0067-0049/192/1/1} {\bibfield  {journal} {\bibinfo
  {journal} {The Astrophysical Journal Supplement Series}\ }\textbf {\bibinfo
  {volume} {192}},\ \bibinfo {pages} {1} (\bibinfo {year} {2010})}\BibitemShut
  {NoStop}%
\bibitem [{\citenamefont {Jimenez}\ and\ \citenamefont
  {Loeb}(2002)}]{Jimenez:2001gg}%
  \BibitemOpen
  \bibfield  {author} {\bibinfo {author} {\bibfnamefont {R.}~\bibnamefont
  {Jimenez}}\ and\ \bibinfo {author} {\bibfnamefont {A.}~\bibnamefont {Loeb}},\
  }\bibfield  {title} {\bibinfo {title} {{Constraining cosmological parameters
  based on relative galaxy ages}},\ }\href {https://doi.org/10.1086/340549}
  {\bibfield  {journal} {\bibinfo  {journal} {Astrophys. J.}\ }\textbf
  {\bibinfo {volume} {573}},\ \bibinfo {pages} {37} (\bibinfo {year} {2002})},\
  \Eprint {https://arxiv.org/abs/astro-ph/0106145} {arXiv:astro-ph/0106145}
  \BibitemShut {NoStop}%
\bibitem [{\citenamefont {Simon}\ \emph {et~al.}(2005)\citenamefont {Simon},
  \citenamefont {Verde},\ and\ \citenamefont {Jimenez}}]{Simon_2005}%
  \BibitemOpen
  \bibfield  {author} {\bibinfo {author} {\bibfnamefont {J.}~\bibnamefont
  {Simon}}, \bibinfo {author} {\bibfnamefont {L.}~\bibnamefont {Verde}},\ and\
  \bibinfo {author} {\bibfnamefont {R.}~\bibnamefont {Jimenez}},\ }\bibfield
  {journal} {\bibinfo  {journal} {Physical Review D}\ }\textbf {\bibinfo
  {volume} {71}},\ \href {https://doi.org/10.1103/physrevd.71.123001}
  {10.1103/physrevd.71.123001} (\bibinfo {year} {2005})\BibitemShut {NoStop}%
\bibitem [{\citenamefont {Stern}\ \emph {et~al.}(2010)\citenamefont {Stern},
  \citenamefont {Jimenez}, \citenamefont {Verde}, \citenamefont
  {Kamionkowski},\ and\ \citenamefont {Stanford}}]{Stern_2010}%
  \BibitemOpen
  \bibfield  {author} {\bibinfo {author} {\bibfnamefont {D.}~\bibnamefont
  {Stern}}, \bibinfo {author} {\bibfnamefont {R.}~\bibnamefont {Jimenez}},
  \bibinfo {author} {\bibfnamefont {L.}~\bibnamefont {Verde}}, \bibinfo
  {author} {\bibfnamefont {M.}~\bibnamefont {Kamionkowski}},\ and\ \bibinfo
  {author} {\bibfnamefont {S.~A.}\ \bibnamefont {Stanford}},\ }\href
  {https://doi.org/10.1088/1475-7516/2010/02/008} {\bibfield  {journal}
  {\bibinfo  {journal} {Journal of Cosmology and Astroparticle Physics}\
  }\textbf {\bibinfo {volume} {2010}}\bibinfo  {number} { (02)},\ \bibinfo
  {pages} {008–008}}\BibitemShut {NoStop}%
\bibitem [{\citenamefont {Wei}(2018)}]{Wei:2018cov}%
  \BibitemOpen
\bibfield  {number} {  }\bibfield  {author} {\bibinfo {author} {\bibfnamefont
  {J.-J.}\ \bibnamefont {Wei}},\ }\bibfield  {title} {\bibinfo {title}
  {{Model-independent Curvature Determination from Gravitational-Wave Standard
  Sirens and Cosmic Chronometers}},\ }\href
  {https://doi.org/10.3847/1538-4357/aae696} {\bibfield  {journal} {\bibinfo
  {journal} {Astrophys. J.}\ }\textbf {\bibinfo {volume} {868}},\ \bibinfo
  {pages} {29} (\bibinfo {year} {2018})},\ \Eprint
  {https://arxiv.org/abs/1806.09781} {arXiv:1806.09781 [astro-ph.CO]}
  \BibitemShut {NoStop}%
\bibitem [{\citenamefont {Blake}\ \emph {et~al.}(2012)\citenamefont {Blake},
  \citenamefont {Brough}, \citenamefont {Colless}, \citenamefont {Contreras},
  \citenamefont {Couch}, \citenamefont {Croom}, \citenamefont {Croton},
  \citenamefont {Davis}, \citenamefont {Drinkwater}, \citenamefont {Forster},\
  and\ \citenamefont {et~al.}}]{Blake_2012}%
  \BibitemOpen
  \bibfield  {author} {\bibinfo {author} {\bibfnamefont {C.}~\bibnamefont
  {Blake}}, \bibinfo {author} {\bibfnamefont {S.}~\bibnamefont {Brough}},
  \bibinfo {author} {\bibfnamefont {M.}~\bibnamefont {Colless}}, \bibinfo
  {author} {\bibfnamefont {C.}~\bibnamefont {Contreras}}, \bibinfo {author}
  {\bibfnamefont {W.}~\bibnamefont {Couch}}, \bibinfo {author} {\bibfnamefont
  {S.}~\bibnamefont {Croom}}, \bibinfo {author} {\bibfnamefont
  {D.}~\bibnamefont {Croton}}, \bibinfo {author} {\bibfnamefont {T.~M.}\
  \bibnamefont {Davis}}, \bibinfo {author} {\bibfnamefont {M.~J.}\ \bibnamefont
  {Drinkwater}}, \bibinfo {author} {\bibfnamefont {K.}~\bibnamefont
  {Forster}},\ and\ \bibinfo {author} {\bibnamefont {et~al.}},\ }\href
  {https://doi.org/10.1111/j.1365-2966.2012.21473.x} {\bibfield  {journal}
  {\bibinfo  {journal} {Monthly Notices of the Royal Astronomical Society}\
  }\textbf {\bibinfo {volume} {425}},\ \bibinfo {pages} {405–414} (\bibinfo
  {year} {2012})}\BibitemShut {NoStop}%
\bibitem [{\citenamefont {Chuang}\ \emph {et~al.}(2013)\citenamefont {Chuang},
  \citenamefont {Prada}, \citenamefont {Cuesta}, \citenamefont {Eisenstein},
  \citenamefont {Kazin}, \citenamefont {Padmanabhan}, \citenamefont {Sanchez},
  \citenamefont {Xu}, \citenamefont {Beutler}, \citenamefont {Manera},\ and\
  \citenamefont {et~al.}}]{Chuang_2013}%
  \BibitemOpen
  \bibfield  {author} {\bibinfo {author} {\bibfnamefont {C.-H.}\ \bibnamefont
  {Chuang}}, \bibinfo {author} {\bibfnamefont {F.}~\bibnamefont {Prada}},
  \bibinfo {author} {\bibfnamefont {A.~J.}\ \bibnamefont {Cuesta}}, \bibinfo
  {author} {\bibfnamefont {D.~J.}\ \bibnamefont {Eisenstein}}, \bibinfo
  {author} {\bibfnamefont {E.}~\bibnamefont {Kazin}}, \bibinfo {author}
  {\bibfnamefont {N.}~\bibnamefont {Padmanabhan}}, \bibinfo {author}
  {\bibfnamefont {A.~G.}\ \bibnamefont {Sanchez}}, \bibinfo {author}
  {\bibfnamefont {X.}~\bibnamefont {Xu}}, \bibinfo {author} {\bibfnamefont
  {F.}~\bibnamefont {Beutler}}, \bibinfo {author} {\bibfnamefont
  {M.}~\bibnamefont {Manera}},\ and\ \bibinfo {author} {\bibnamefont
  {et~al.}},\ }\href {https://doi.org/10.1093/mnras/stt988} {\bibfield
  {journal} {\bibinfo  {journal} {Monthly Notices of the Royal Astronomical
  Society}\ }\textbf {\bibinfo {volume} {433}},\ \bibinfo {pages} {3559–3571}
  (\bibinfo {year} {2013})}\BibitemShut {NoStop}%
\bibitem [{\citenamefont {Font-Ribera}\ \emph {et~al.}(2014)\citenamefont
  {Font-Ribera}, \citenamefont {Kirkby}, \citenamefont {Busca}, \citenamefont
  {Miralda-Escudé}, \citenamefont {Ross}, \citenamefont {Slosar},
  \citenamefont {Rich}, \citenamefont {Aubourg}, \citenamefont {Bailey},
  \citenamefont {Bhardwaj},\ and\ \citenamefont {et~al.}}]{Font_Ribera_2014}%
  \BibitemOpen
  \bibfield  {author} {\bibinfo {author} {\bibfnamefont {A.}~\bibnamefont
  {Font-Ribera}}, \bibinfo {author} {\bibfnamefont {D.}~\bibnamefont {Kirkby}},
  \bibinfo {author} {\bibfnamefont {N.}~\bibnamefont {Busca}}, \bibinfo
  {author} {\bibfnamefont {J.}~\bibnamefont {Miralda-Escudé}}, \bibinfo
  {author} {\bibfnamefont {N.~P.}\ \bibnamefont {Ross}}, \bibinfo {author}
  {\bibfnamefont {A.}~\bibnamefont {Slosar}}, \bibinfo {author} {\bibfnamefont
  {J.}~\bibnamefont {Rich}}, \bibinfo {author} {\bibfnamefont {Ã.}~\bibnamefont
  {Aubourg}}, \bibinfo {author} {\bibfnamefont {S.}~\bibnamefont {Bailey}},
  \bibinfo {author} {\bibfnamefont {V.}~\bibnamefont {Bhardwaj}},\ and\
  \bibinfo {author} {\bibnamefont {et~al.}},\ }\href
  {https://doi.org/10.1088/1475-7516/2014/05/027} {\bibfield  {journal}
  {\bibinfo  {journal} {Journal of Cosmology and Astroparticle Physics}\
  }\textbf {\bibinfo {volume} {2014}}\bibinfo  {number} { (05)},\ \bibinfo
  {pages} {027–027}}\BibitemShut {NoStop}%
\bibitem [{\citenamefont {Delubac}\ \emph {et~al.}(2015)\citenamefont
  {Delubac}, \citenamefont {Bautista}, \citenamefont {Busca}, \citenamefont
  {Rich}, \citenamefont {Kirkby}, \citenamefont {Bailey}, \citenamefont
  {Font-Ribera}, \citenamefont {Slosar}, \citenamefont {Lee}, \citenamefont
  {Pieri},\ and\ \citenamefont {et~al.}}]{Delubac_2015}%
  \BibitemOpen
\bibfield  {number} {  }\bibfield  {author} {\bibinfo {author} {\bibfnamefont
  {T.}~\bibnamefont {Delubac}}, \bibinfo {author} {\bibfnamefont {J.~E.}\
  \bibnamefont {Bautista}}, \bibinfo {author} {\bibfnamefont {N.~G.}\
  \bibnamefont {Busca}}, \bibinfo {author} {\bibfnamefont {J.}~\bibnamefont
  {Rich}}, \bibinfo {author} {\bibfnamefont {D.}~\bibnamefont {Kirkby}},
  \bibinfo {author} {\bibfnamefont {S.}~\bibnamefont {Bailey}}, \bibinfo
  {author} {\bibfnamefont {A.}~\bibnamefont {Font-Ribera}}, \bibinfo {author}
  {\bibfnamefont {A.}~\bibnamefont {Slosar}}, \bibinfo {author} {\bibfnamefont
  {K.-G.}\ \bibnamefont {Lee}}, \bibinfo {author} {\bibfnamefont {M.~M.}\
  \bibnamefont {Pieri}},\ and\ \bibinfo {author} {\bibnamefont {et~al.}},\
  }\href {https://doi.org/10.1051/0004-6361/201423969} {\bibfield  {journal}
  {\bibinfo  {journal} {Astronomy \& Astrophysics}\ }\textbf {\bibinfo {volume}
  {574}},\ \bibinfo {pages} {A59} (\bibinfo {year} {2015})}\BibitemShut
  {NoStop}%
\bibitem [{\citenamefont {Bautista}\ \emph {et~al.}(2017)\citenamefont
  {Bautista}, \citenamefont {Busca}, \citenamefont {Guy}, \citenamefont {Rich},
  \citenamefont {Blomqvist}, \citenamefont {du~Mas~des Bourboux}, \citenamefont
  {Pieri}, \citenamefont {Font-Ribera}, \citenamefont {Bailey}, \citenamefont
  {Delubac},\ and\ \citenamefont {et~al.}}]{Bautista_2017}%
  \BibitemOpen
  \bibfield  {author} {\bibinfo {author} {\bibfnamefont {J.~E.}\ \bibnamefont
  {Bautista}}, \bibinfo {author} {\bibfnamefont {N.~G.}\ \bibnamefont {Busca}},
  \bibinfo {author} {\bibfnamefont {J.}~\bibnamefont {Guy}}, \bibinfo {author}
  {\bibfnamefont {J.}~\bibnamefont {Rich}}, \bibinfo {author} {\bibfnamefont
  {M.}~\bibnamefont {Blomqvist}}, \bibinfo {author} {\bibfnamefont
  {H.}~\bibnamefont {du~Mas~des Bourboux}}, \bibinfo {author} {\bibfnamefont
  {M.~M.}\ \bibnamefont {Pieri}}, \bibinfo {author} {\bibfnamefont
  {A.}~\bibnamefont {Font-Ribera}}, \bibinfo {author} {\bibfnamefont
  {S.}~\bibnamefont {Bailey}}, \bibinfo {author} {\bibfnamefont
  {T.}~\bibnamefont {Delubac}},\ and\ \bibinfo {author} {\bibnamefont
  {et~al.}},\ }\href {https://doi.org/10.1051/0004-6361/201730533} {\bibfield
  {journal} {\bibinfo  {journal} {Astronomy \& Astrophysics}\ }\textbf
  {\bibinfo {volume} {603}},\ \bibinfo {pages} {A12} (\bibinfo {year}
  {2017})}\BibitemShut {NoStop}%
\bibitem [{\citenamefont {Foreman-Mackey}\ \emph {et~al.}(2013)\citenamefont
  {Foreman-Mackey}, \citenamefont {Hogg}, \citenamefont {Lang},\ and\
  \citenamefont {Goodman}}]{Foreman_Mackey_2013}%
  \BibitemOpen
  \bibfield  {author} {\bibinfo {author} {\bibfnamefont {D.}~\bibnamefont
  {Foreman-Mackey}}, \bibinfo {author} {\bibfnamefont {D.~W.}\ \bibnamefont
  {Hogg}}, \bibinfo {author} {\bibfnamefont {D.}~\bibnamefont {Lang}},\ and\
  \bibinfo {author} {\bibfnamefont {J.}~\bibnamefont {Goodman}},\ }\href
  {https://doi.org/10.1086/670067} {\bibfield  {journal} {\bibinfo  {journal}
  {Publications of the Astronomical Society of the Pacific}\ }\textbf {\bibinfo
  {volume} {125}},\ \bibinfo {pages} {306–312} (\bibinfo {year}
  {2013})}\BibitemShut {NoStop}%
\bibitem [{\citenamefont {{Goodman}}\ and\ \citenamefont
  {{Weare}}(2010)}]{2010CAMCS...5...65G}%
  \BibitemOpen
  \bibfield  {author} {\bibinfo {author} {\bibfnamefont {J.}~\bibnamefont
  {{Goodman}}}\ and\ \bibinfo {author} {\bibfnamefont {J.}~\bibnamefont
  {{Weare}}},\ }\href {https://doi.org/10.2140/camcos.2010.5.65} {\bibfield
  {journal} {\bibinfo  {journal} {Communications in Applied Mathematics and
  Computational Science}\ }\textbf {\bibinfo {volume} {5}},\ \bibinfo {pages}
  {65} (\bibinfo {year} {2010})}\BibitemShut {NoStop}%
\bibitem [{\citenamefont {Lewis}(2019)}]{Lewis:2019xzd}%
  \BibitemOpen
  \bibfield  {author} {\bibinfo {author} {\bibfnamefont {A.}~\bibnamefont
  {Lewis}},\ }\bibfield  {title} {\bibinfo {title} {{GetDist: a Python package
  for analysing Monte Carlo samples}},\ }\href@noop {} {\  (\bibinfo {year}
  {2019})},\ \Eprint {https://arxiv.org/abs/1910.13970} {arXiv:1910.13970
  [astro-ph.IM]} \BibitemShut {NoStop}%
\bibitem [{\citenamefont {Group}(2020)}]{10.1093/ptep/ptaa104}%
  \BibitemOpen
  \bibfield  {author} {\bibinfo {author} {\bibfnamefont {P.~D.}\ \bibnamefont
  {Group}},\ }\bibfield  {title} {\bibinfo {title} {{Review of Particle
  Physics}},\ }\bibfield  {journal} {\bibinfo  {journal} {Progress of
  Theoretical and Experimental Physics}\ }\textbf {\bibinfo {volume} {2020}},\
  \href {https://doi.org/10.1093/ptep/ptaa104} {10.1093/ptep/ptaa104} (\bibinfo
  {year} {2020}),\ \bibinfo {note} {083C01},\ \Eprint
  {https://arxiv.org/abs/https://academic.oup.com/ptep/article-pdf/2020/8/083C01/33653179/ptaa104.pdf}
  {https://academic.oup.com/ptep/article-pdf/2020/8/083C01/33653179/ptaa104.pdf}
  \BibitemShut {NoStop}%
\bibitem [{\citenamefont {Aghanim}\ \emph {et~al.}(2020)\citenamefont {Aghanim}
  \emph {et~al.}}]{Aghanim:2018eyx}%
  \BibitemOpen
  \bibfield  {author} {\bibinfo {author} {\bibfnamefont {N.}~\bibnamefont
  {Aghanim}} \emph {et~al.} (\bibinfo {collaboration} {Planck}),\ }\bibfield
  {title} {\bibinfo {title} {{Planck 2018 results. VI. Cosmological
  parameters}},\ }\href {https://doi.org/10.1051/0004-6361/201833910}
  {\bibfield  {journal} {\bibinfo  {journal} {Astron. Astrophys.}\ }\textbf
  {\bibinfo {volume} {641}},\ \bibinfo {pages} {A6} (\bibinfo {year} {2020})},\
  \Eprint {https://arxiv.org/abs/1807.06209} {arXiv:1807.06209 [astro-ph.CO]}
  \BibitemShut {NoStop}%
\bibitem [{\citenamefont {Riess}\ \emph {et~al.}(2019)\citenamefont {Riess},
  \citenamefont {Casertano}, \citenamefont {Yuan}, \citenamefont {Macri},\ and\
  \citenamefont {Scolnic}}]{Riess:2019cxk}%
  \BibitemOpen
  \bibfield  {author} {\bibinfo {author} {\bibfnamefont {A.~G.}\ \bibnamefont
  {Riess}}, \bibinfo {author} {\bibfnamefont {S.}~\bibnamefont {Casertano}},
  \bibinfo {author} {\bibfnamefont {W.}~\bibnamefont {Yuan}}, \bibinfo {author}
  {\bibfnamefont {L.~M.}\ \bibnamefont {Macri}},\ and\ \bibinfo {author}
  {\bibfnamefont {D.}~\bibnamefont {Scolnic}},\ }\bibfield  {title} {\bibinfo
  {title} {{Large Magellanic Cloud Cepheid Standards Provide a 1\% Foundation
  for the Determination of the Hubble Constant and Stronger Evidence for
  Physics beyond $\Lambda$CDM}},\ }\href
  {https://doi.org/10.3847/1538-4357/ab1422} {\bibfield  {journal} {\bibinfo
  {journal} {Astrophys. J.}\ }\textbf {\bibinfo {volume} {876}},\ \bibinfo
  {pages} {85} (\bibinfo {year} {2019})},\ \Eprint
  {https://arxiv.org/abs/1903.07603} {arXiv:1903.07603 [astro-ph.CO]}
  \BibitemShut {NoStop}%
\bibitem [{\citenamefont {Trotta}(2008)}]{Trotta:2008qt}%
  \BibitemOpen
  \bibfield  {author} {\bibinfo {author} {\bibfnamefont {R.}~\bibnamefont
  {Trotta}},\ }\bibfield  {title} {\bibinfo {title} {{Bayes in the sky:
  Bayesian inference and model selection in cosmology}},\ }\href
  {https://doi.org/10.1080/00107510802066753} {\bibfield  {journal} {\bibinfo
  {journal} {Contemp. Phys.}\ }\textbf {\bibinfo {volume} {49}},\ \bibinfo
  {pages} {71} (\bibinfo {year} {2008})},\ \Eprint
  {https://arxiv.org/abs/0803.4089} {arXiv:0803.4089 [astro-ph]} \BibitemShut
  {NoStop}%
\bibitem [{\citenamefont {Liddle}(2007)}]{Liddle:2007fy}%
  \BibitemOpen
  \bibfield  {author} {\bibinfo {author} {\bibfnamefont {A.~R.}\ \bibnamefont
  {Liddle}},\ }\bibfield  {title} {\bibinfo {title} {{Information criteria for
  astrophysical model selection}},\ }\href
  {https://doi.org/10.1111/j.1745-3933.2007.00306.x} {\bibfield  {journal}
  {\bibinfo  {journal} {Mon. Not. Roy. Astron. Soc.}\ }\textbf {\bibinfo
  {volume} {377}},\ \bibinfo {pages} {L74} (\bibinfo {year} {2007})},\ \Eprint
  {https://arxiv.org/abs/astro-ph/0701113} {arXiv:astro-ph/0701113}
  \BibitemShut {NoStop}%
\bibitem [{\citenamefont {Wei}(2010)}]{Hwei}%
  \BibitemOpen
  \bibfield  {author} {\bibinfo {author} {\bibfnamefont {H.}~\bibnamefont
  {Wei}},\ }\bibfield  {title} {\bibinfo {title} {Revisiting the cosmological
  constraints on the interacting dark energy models},\ }\href
  {https://doi.org/10.1016/j.physletb.2010.06.038} {\bibfield  {journal}
  {\bibinfo  {journal} {Physics Letters B}\ }\textbf {\bibinfo {volume}
  {691}},\ \bibinfo {pages} {173–182} (\bibinfo {year} {2010})}\BibitemShut
  {NoStop}%
\bibitem [{\citenamefont {Rosenfeld}(2007)}]{RRos}%
  \BibitemOpen
  \bibfield  {author} {\bibinfo {author} {\bibfnamefont {R.}~\bibnamefont
  {Rosenfeld}},\ }\bibfield  {title} {\bibinfo {title} {Reconstruction of
  interacting dark energy models from parametrizations},\ }\bibfield  {journal}
  {\bibinfo  {journal} {Physical Review D}\ }\textbf {\bibinfo {volume} {75}},\
  \href {https://doi.org/10.1103/physrevd.75.083509}
  {10.1103/physrevd.75.083509} (\bibinfo {year} {2007})\BibitemShut {NoStop}%
\bibitem [{\citenamefont {Yang}\ \emph {et~al.}(2020)\citenamefont {Yang},
  \citenamefont {Di~Valentino}, \citenamefont {Mena}, \citenamefont {Pan},\
  and\ \citenamefont {Nunes}}]{Yang:2020uga}%
  \BibitemOpen
  \bibfield  {author} {\bibinfo {author} {\bibfnamefont {W.}~\bibnamefont
  {Yang}}, \bibinfo {author} {\bibfnamefont {E.}~\bibnamefont {Di~Valentino}},
  \bibinfo {author} {\bibfnamefont {O.}~\bibnamefont {Mena}}, \bibinfo {author}
  {\bibfnamefont {S.}~\bibnamefont {Pan}},\ and\ \bibinfo {author}
  {\bibfnamefont {R.~C.}\ \bibnamefont {Nunes}},\ }\bibfield  {title} {\bibinfo
  {title} {{All-inclusive interacting dark sector cosmologies}},\ }\href
  {https://doi.org/10.1103/PhysRevD.101.083509} {\bibfield  {journal} {\bibinfo
   {journal} {Phys. Rev. D}\ }\textbf {\bibinfo {volume} {101}},\ \bibinfo
  {pages} {083509} (\bibinfo {year} {2020})},\ \Eprint
  {https://arxiv.org/abs/2001.10852} {arXiv:2001.10852 [astro-ph.CO]}
  \BibitemShut {NoStop}%
\end{thebibliography}%

\end{document}